\documentclass[11pt]{article}
\usepackage[utf8]{inputenc}
\usepackage[T1]{fontenc}
\usepackage{lmodern}
\usepackage{hyperref}
\usepackage{url}
\usepackage{natbib}
\usepackage{booktabs}
\usepackage{xcolor}
\usepackage{listings}
\usepackage[margin=1in]{geometry}
\usepackage{graphicx}
\usepackage{amsmath}
\usepackage{orcidlink}
\usepackage{enumitem}
\usepackage{tikz}
\usetikzlibrary{arrows.meta, positioning}

\title{scicode-lint: Detecting Methodology Bugs in Scientific Python Code with LLM-Generated Patterns}
\author{
    Sergey V. Samsonau\,\orcidlink{0000-0002-0835-2970}\thanks{Corresponding author: \href{mailto:ssamsonau@gradcenter.cuny.edu}{ssamsonau@gradcenter.cuny.edu}}\\[6pt]
    Authentic Research Partners, Princeton, NJ\thanks{\url{https://arpconnect.com/}}
}
\date{}

\begin{document}
\pagestyle{plain}
\maketitle

\begin{abstract}
Methodology bugs in scientific Python code produce plausible but incorrect results that traditional linters and static analysis tools cannot detect. Several research groups have built ML-specific linters, demonstrating that detection is feasible. Yet these tools share a sustainability problem: dependency on specific pylint or Python versions, limited packaging, and reliance on manual engineering for every new pattern. As AI-generated code increases the volume of scientific software, the need for automated methodology checking (such as detecting data leakage, incorrect cross-validation, and missing random seeds) grows. We present \textbf{scicode-lint}\footnote{Code: \url{https://github.com/ssamsonau/scicode-lint}. PyPI: \url{https://pypi.org/project/scicode-lint/} (v0.4.0).}, whose two-tier architecture separates pattern design (frontier models at build time) from execution (small local model at runtime). Patterns are generated, not hand-coded; adapting to new library versions costs tokens, not engineering hours. On Kaggle notebooks with human-labeled ground truth, preprocessing leakage detection reaches 65\% precision at 100\% recall; on 38 published scientific papers applying AI/ML, precision is 62\% (LLM-judged) with substantial variation across pattern categories; on a held-out paper set, precision is 54\%. On controlled tests, scicode-lint achieves 97.7\% accuracy across 66 patterns.
\end{abstract}

\section{Introduction}

Methodology bugs in scientific Python code do not cause crashes. They produce plausible but incorrect results. Traditional linters cannot detect them; these bugs require semantic understanding of scientific methodology. For example, scicode-lint found a published vocoder repository that accepts a \texttt{--random-state} command-line argument but never calls \texttt{np.random.seed()} or \texttt{random.seed()} with it. Data splitting is non-reproducible despite the explicit parameter; a researcher re-running the experiment with the same seed would get different results without any indication of why. This is one of 66 patterns the tool detects across five categories.

Several groups have addressed this: dslinter~\citep{haakman2022dslinter}, MLScent~\citep{shivashankar2025mlscent}, mllint~\citep{vanoort2022mllint}, among others. These tools demonstrated feasibility but face recurring sustainability challenges (Section~\ref{sec:tools}).

We present scicode-lint, a linter that uses a local LLM to detect methodology bugs in scientific Python code. Its two-tier architecture (Section~\ref{sec:arch}) separates pattern design (frontier models at build time) from execution (small local model at runtime on commodity hardware). Adapting to changes costs tokens, not engineering hours. Contributors develop evaluation harnesses and quality gates rather than writing detection rules by hand.

\section{Background and Related Work}

\subsection{Scientific Research Runs on Code}

Modern scientific research depends on software at every stage, from data collection through analysis to publication. Over half of researchers develop their own software~\citep{hettrick2014survey}, and AI/ML have become a major component. The demand for reliable scientific software has driven the growth of Research Software Engineering (RSE) as a field, yet the estimated 10,000 RSEs worldwide serve a research community of millions; most research code is written by scientists for whom software development is a means to an end, not a core competency, and who rarely have access to expert code review.

\subsection{Bugs, Reproducibility, and Data Leakage}

Scientific code has always harbored bugs, but the most dangerous class is semantic: errors that produce plausible but wrong results. Hatton's T experiments~\citep{hatton1997t} found that independent implementations of the same algorithms progressively degenerated from 6 to 1 significant figures (i.e., independent implementations of identical algorithms diverged by orders of magnitude). \citet{soergel2015rampant} estimated that for a medium-scale bioinformatics analysis, the probability of at least one output-altering error is effectively 100\%. The Reinhart-Rogoff Excel error (an AVERAGE formula omitting 5 countries) inverted an economic conclusion that influenced austerity policies worldwide~\citep{herndon2014debt}. The reproducibility crisis confirms this is not anecdotal: \citet{stodden2018empirical} found a 26\% computational reproducibility rate in \textit{Science} articles.

As scientific code increasingly involves ML, the problem takes a specific form: methodology bugs such as data leakage, incorrect cross-validation, and missing random seeds. These are particularly insidious because they inflate performance metrics rather than producing obviously wrong output. \citet{kapoor2023leakage} documented leakage errors affecting 329 papers across 17 fields. \citet{roberts2021covid} reviewed 2,212 COVID-19 ML studies and found none of the 61 that passed quality screening clinically useful. In neuroimaging, \citet{vandemortel2025leakage} found 45\% of MRI studies in a meta-analysis had procedures consistent with leakage. Each research community tends to rediscover these pitfalls independently, suggesting the problem is structural~\citep{kapoor2023leakage}.

\subsection{AI-Generated Code}

``Vibe coding''\footnote{Term coined by A. Karpathy, post on X, February 2025: \url{https://x.com/karpathy/status/1886192184808149383}.}, the practice of accepting AI-generated code with minimal review, adds another dimension. In a controlled study, developers using an AI coding assistant wrote significantly less secure code while being more likely to believe it was secure~\citep{perry2023insecure}. A survey of 868 scientists who program found that most report inadequate training for the software development their work demands, alongside limited use of practices such as testing and code review~\citep{obrien2025survey}.

The industry response has been swift: automated quality tools that run after AI-generated changes (GitHub Copilot code review, Aider's linter-in-the-loop). These tools catch general software bugs: logic errors, security vulnerabilities, style violations. What they do not catch are methodology bugs specific to scientific code. scicode-lint brings the linter-in-the-loop principle to this domain.

\subsection{Existing Code Quality Tools}
\label{sec:tools}

\textbf{ML-specific linters.} dslinter~\citep{haakman2022dslinter} provides $\sim$25 AST-based checkers but publishes no accuracy benchmarks. MLScent~\citep{shivashankar2025mlscent} implements 76 detectors (the most comprehensive ML-specific static analysis) but was never released on PyPI. mllint~\citep{vanoort2022mllint} checks project hygiene (e.g., use of version control, dependency management, CI) rather than code-level bug detection. DyLin~\citep{eghbali2025dylin}, the first dynamic Python linter, achieved 70.6\% precision but only 3 of its 15 checkers touch ML.

For patterns expressible as AST or data-flow rules, traditional static analysis is currently more reliable in practice. But these tools share sustainability problems: mllint's author announced the project is no longer maintained after graduating~\citep{vanoort2022mllint}; dslinter's last release predates the pylint version that broke it; MLScent was never packaged. This reflects a structural under-investment in research software engineering~\citep{deschamps2023better}, not individual failing.

\begin{table}[ht]
\centering
\vspace{0.5em}
\caption{ML and scientific code linting tools (March 2026).}
\label{tab:tools}
\small
\begin{tabular}{@{}llp{5.5cm}@{}}
\toprule
\textbf{Tool} & \textbf{Patterns} & \textbf{Status} \\
\midrule
dslinter~\citep{haakman2022dslinter} & $\sim$25 & PyPI, Jun 2022; broken by pylint 3.0 \\ \midrule
MLScent~\citep{shivashankar2025mlscent} & 76 & Not on PyPI \\ \midrule
mllint~\citep{vanoort2022mllint} & N/A & ``No longer maintained'' \\ \midrule
DyLin~\citep{eghbali2025dylin} & 15 (3 ML) & Not on PyPI; requires Docker \\ \midrule
\citet{yang2022leakage} & 3 (leakage) & GitHub only; 92.9\% validation acc. \\ \midrule
\textbf{scicode-lint} & \textbf{66} & \textbf{PyPI + GitHub; 97.7\% controlled; 65\% P, 100\% R on labeled notebooks} \\
\bottomrule
\end{tabular}
\end{table}

Based on documentation review, existing tools cover approximately 9--14\% of scicode-lint's 66 patterns. No existing tool documents checks for PyTorch inference mode errors, temporal leakage, target encoding leakage, or numerical stability patterns, categories accounting for over 50 of 66 patterns.

\textbf{Static leakage detection.} \citet{yang2022leakage} developed static data-flow analysis for 3 leakage types, achieving 92.9\% accuracy on their validation set and applying the tool to over 100K Kaggle notebooks; \citet{drobnjakovic2024abstract} achieved 93\% precision using abstract interpretation. For their targeted leakage types, these tools currently achieve higher precision. Their limitation is the range of patterns covered and the difficulty of extension.

\textbf{Hybrid LLM-plus-static-analysis.} LLift~\citep{li2024lllift} and KNighter~\citep{yang2025knighter} combine LLMs with static analysis for security vulnerabilities in systems code, validating the hybrid paradigm. scicode-lint applies the same principle to scientific methodology with the additional constraint of local execution on commodity hardware.

\section{Architecture}
\label{sec:arch}

\subsection{Two-Tier Strategy}

The core insight: designing good detection questions is hard, but executing them is easy.

At build time, frontier models design detection patterns: analyzing library documentation, designing focused detection questions, and generating test cases. In our implementation, Claude Opus 4.6 (via Claude Code) handles interactive development, while Claude Sonnet 4.6 handles all automated tasks: semantic validation, integration test generation and judging, and real-world finding verification.

At runtime, a small local model (RedHatAI/Qwen3-8B-FP8-dynamic) executes pre-designed detection questions against user code on an NVIDIA RTX 4000 Ada (20GB VRAM). The model receives a focused question, the code, and returns a structured verdict. vLLM~\citep{kwon2023vllm} serves the model with PagedAttention and prefix caching.

This design invests expensive reasoning once and amortizes across unlimited local executions. It also addresses sustainability: adapting to new library versions requires re-running the build-time step against updated documentation. The cost is measured in tokens, not engineering hours. When the build-time model improves, patterns improve without infrastructure changes. The build-time step is not tied to any specific provider.

\subsection{Pattern Structure}

Each pattern targets a specific anti-pattern and contains: a detection question (single-issue, self-contained, with context on why the issue matters), references to official documentation, and test files (3+ positive, 3+ negative). We maintain 66 patterns across 5 categories: AI training (19), AI inference (12), scientific-numerical (10), scientific-performance (11), and scientific-reproducibility (14). Users can select relevant categories to reduce scan time and noise.

\textbf{Example: Pattern ml-001 (scaler leakage).} The detection question is a self-contained prompt that the runtime model receives alongside the user's code:

\noindent\textit{Detection question (abbreviated):}
\begin{quote}\small
Analyze data preprocessing for data leakage. If preprocessing statistics (mean, std, min, max) are computed on the full dataset including test data, then test set information leaks into training. Look for: scaler fit\_transform() called on full data BEFORE train\_test\_split(); train and test arrays concatenated BEFORE computing statistics. NOT a bug: statistics loaded from constants or config files; code with comments stating this is intentional; statistics computed on training data only, after split. YES = preprocessing statistics computed from data that includes test samples. NO = statistics computed on training data only, or loaded from constants.
\end{quote}

\begin{quote}
\noindent\textit{Positive test (bug present):}
\begin{lstlisting}
def normalize_with_test_stats(train_data, test_data):
    full_dataset = torch.vstack([train_data, test_data])
    dataset_mean = full_dataset.mean()
    dataset_std = full_dataset.std()
    train_normalized = (train_data - dataset_mean) / dataset_std
    test_normalized = (test_data - dataset_mean) / dataset_std
    return train_normalized, test_normalized
\end{lstlisting}

\noindent\textit{Negative test (correct code):}
\begin{lstlisting}
def normalize_correctly(train_data, test_data):
    mean = train_data.mean()
    std = train_data.std()
    train_normalized = (train_data - mean) / std
    test_normalized = (test_data - mean) / std
    return train_normalized, test_normalized
\end{lstlisting}
\end{quote}

The detection question includes explicit YES/NO framing and false-positive exclusions (e.g., statistics loaded from constants) that help the runtime model distinguish genuine bugs from intentional patterns.

\subsection{Performance and Security}

\textbf{Caching and throughput.} Code-first prompt ordering (the user's code appears before the detection question) means all 66 prompts for a given file share the same prefix; vLLM caches this shared prefix in the KV cache (hit rates regularly exceeding 75\%); all 66 patterns run concurrently via async batching, with the shared prefix computed only once. vLLM's structured-output mode (\texttt{response\_format}) constrains the model's output to a predefined JSON schema while preserving Qwen3's chain-of-thought reasoning phase; Pydantic validates the parsed response on the client side.

\textbf{Timing.} A full scan of one file against all 66 patterns runs in $\sim$1 minute for small files and $\sim$2.5 minutes for files approaching 1,000 lines on a single RTX 4000 Ada GPU. Scan time scales sub-linearly with file size: a 33$\times$ increase in lines (30 to 984) adds only $\sim$2.5$\times$ to the time, because all 66 patterns share the cached code prefix and run concurrently rather than as 66 independent scans. Files exceeding the model's input budget are skipped with a reported reason, not silently dropped.

\textbf{Prompt injection defense:} three-layer defense-in-depth (system message framing, XML delimiters, post-code reinforcement).

\textbf{Dual-audience output:} structured JSON for AI coding agents and human-readable explanations with documentation links. As AI agents become primary authors of scientific code, linter output must be consumable by both agent and researcher.

\textbf{Detection only:} no automatic fixes. The user reviews all findings.

\section{Evaluation}

\subsection{Framework}

\begin{table}[ht]
\centering
\caption{Evaluation framework.}
\label{tab:eval}
\begin{tabular}{@{}llp{5cm}@{}}
\toprule
\textbf{Layer} & \textbf{Optimized?} & \textbf{Key result} \\
\midrule
Pattern evals & Yes & 97.7\% accuracy \\ \midrule
Integration (n=50) & No & 58\% P, 85\% R \\ \midrule
Kaggle labeled~\citep{yang2022leakage} & No & 65\% P, 100\% R (preprocessing leakage) \\ \midrule
PapersWithCode (feedback)~\citep{paperswithcode} & Feedback iter. & 62\% P \\ \midrule
PapersWithCode (holdout)~\citep{paperswithcode} & No & 54\% P \\
\bottomrule
\end{tabular}
\end{table}

Pattern evals test patterns against their own test files. Integration scenarios and Kaggle notebooks were never seen during pattern development. The PapersWithCode evaluation uses two non-overlapping paper sets: a feedback set (38 papers, 119 self-contained files) where verification feedback refined detection questions, and a holdout set (35 papers, 45 self-contained files) sampled after all iteration was complete. The 8-point precision gap (62\% vs.\ 54\%) indicates patterns generalize reasonably, with per-category performance consistent across both sets.

The integration evaluation uses a fully automated four-step pipeline. (1)~Sonnet selects 2--3 patterns that would naturally coexist in one file. (2)~Sonnet generates 30--60 lines of realistic scientific Python containing exactly those bugs, with natural variable names and no revealing comments. (3)~A separate Sonnet call verifies each bug is genuinely present at the claimed location, with up to 3 regeneration attempts. (4)~scicode-lint runs detection and Sonnet judges each finding against the ground-truth manifest. Across 50 scenarios with 148 intended bugs: 85.1\% recall, 58.0\% precision, F1 = 69.1\%. The 27 bonus findings (real bugs beyond what was planted) confirm that the generated code is realistic enough to contain unintended issues.

All results represent a baseline: the smallest viable runtime model (8B parameters), first-generation patterns, single-file analysis, and no model fine-tuning. Every component of the architecture is designed to improve independently: better patterns, larger models, multi-file context, all without requiring changes to the rest of the system.

The Kaggle evaluation is the only layer with human ground truth rather than LLM verification. On Yang et al.'s human-labeled notebooks~\citep{yang2022leakage}, patterns were not tuned against this benchmark. Preprocessing leakage detection achieves 65\% precision at 100\% recall (F1 = 79\%); multi-test leakage reaches 80\% precision at 18\% recall. Overlap leakage remains undetected.

\subsection{Validation on Published Scientific Code}

We ran scicode-lint on code from published scientific papers that apply AI/ML to scientific domains, sourced via PapersWithCode~\citep{paperswithcode} (data collected before the platform's July 2025 shutdown). Papers were selected through multi-stage filtering: domain keyword filters across 8 scientific domains (excluding 29 pure ML venues), LLM classification of abstracts, and domain-balanced sampling. From each paper's repository, Python files with ML imports were identified; a prefilter classified each as a self-contained ML pipeline or a code fragment, retaining only self-contained files for analysis. All findings were verified by a frontier LLM (Claude Sonnet 4.6), not human-labeled ground truth.

\textbf{Verification methodology.} Each finding is verified by Claude Sonnet 4.6 in a separate call with structured reasoning: the judge receives the pattern category, severity, the runtime model's issue description and explanation, the flagged code snippet with surrounding context, and the file's repository origin. It does not receive the detection question itself; the judge sees what was found and the code, not what was asked. The judge produces a chain-of-thought assessment and a verdict (valid, invalid, or uncertain). Precision is computed conservatively: uncertain findings are included in the denominator (i.e., treated as not-valid), giving 62\% rather than 65\% if uncertain were excluded. The Kaggle evaluation (Section~4.1), which uses human-labeled ground truth, provides an independent calibration point.

\textbf{Feedback set.} 38 papers, 119 self-contained files (13\% of 884 qualifying files). Overall precision: 62\% (85 valid, 45 invalid, 7 uncertain).

\begin{table}[ht]
\centering
\caption{Feedback set precision by severity.}
\label{tab:severity}
\begin{tabular}{@{}lrrr@{}}
\toprule
\textbf{Severity} & \textbf{Findings} & \textbf{Valid} & \textbf{Precision} \\
\midrule
Critical & 21 & 5 & 24\% \\ \midrule
High & 77 & 52 & 68\% \\ \midrule
Medium & 39 & 28 & 72\% \\
\bottomrule
\end{tabular}
\end{table}

Critical findings (data leakage, missing zero\_grad()) have lower precision (24\%), reflecting the difficulty of formulating detection questions precise enough for these patterns. High and medium findings (missing map\_location, CUDA non-determinism, loop vectorization) reach 68--72\% precision. Of the 32 papers with self-contained files, 24 (75\%) yielded at least one verified valid finding, across domains from chemistry and earth science to biology and economics. This is consistent with the broader evidence that scientific software frequently harbors errors~\citep{hatton1997t,soergel2015rampant}.

\textbf{Holdout set.} A separate sample of 35 papers, drawn after all pattern refinement was complete and excluding all feedback-set papers, yielded 17 papers with self-contained files (45 files total). Overall precision: 54\% (40 valid, 28 invalid, 6 uncertain out of 74 findings). Of the 17 papers, 12 (71\%) had verified real bugs, consistent with the feedback set (75\%). Performance varied substantially by category: ai-inference (62\% P) and scientific-reproducibility (67\% P) generalized well, while ai-training (11\% P) and scientific-performance (25\% P) showed high false-positive rates, identifying clear targets for pattern improvement.

\textbf{Gap analysis.} The gap between controlled (97.7\%) and real-world (62\%) accuracy reflects the difference between curated test files and real code. Real-world code uses patterns, abstractions, and library-specific idioms that differ substantially from controlled test cases. Filtering to self-contained pipeline files (rather than analyzing all Python files) substantially reduced false positives, confirming that analyzing code fragments out of context is a major source of noise.

\section{Development Process}

The tool was produced using AI-assisted development: the developer writes specifications, AI implements, and layered automated gates enforce quality.

\subsection{Quality Gates}

\begin{table}[ht]
\centering
\caption{Quality gate layers.}
\label{tab:gates}
\begin{tabular}{@{}lp{5.5cm}l@{}}
\toprule
\textbf{Layer} & \textbf{What it checks} & \textbf{Cost} \\
\midrule
Deterministic (15 checks) & Structure, syntax, file sync, data leakage in test files & Free, seconds \\ \midrule
Diversity check & Redundant tests, non-diverse negatives (semantic similarity) & Local vLLM \\ \midrule
Semantic validation & Detection question $\leftrightarrow$ test file $\leftrightarrow$ documentation consistency & Frontier model tokens \\ \midrule
Pattern evals & Detection accuracy (P, R, F1) & Local vLLM \\ \midrule
Integration tests & Generalization to LLM-generated code (50 scenarios) & Local vLLM + frontier \\ \midrule
Real-world validation & Precision on actual scientific code & Local vLLM + frontier \\
\bottomrule
\end{tabular}
\end{table}

Each layer filters different problems. A pattern can be structurally valid, semantically consistent, accurate on its own tests, and still produce false positives on real-world code. This is why all six layers are necessary.

The deterministic layer includes its own data leakage check: test files must not contain hint comments (e.g., \texttt{\# BUG:}) that would leak ground truth to the runtime model.

\subsection{Self-Improvement Loop}

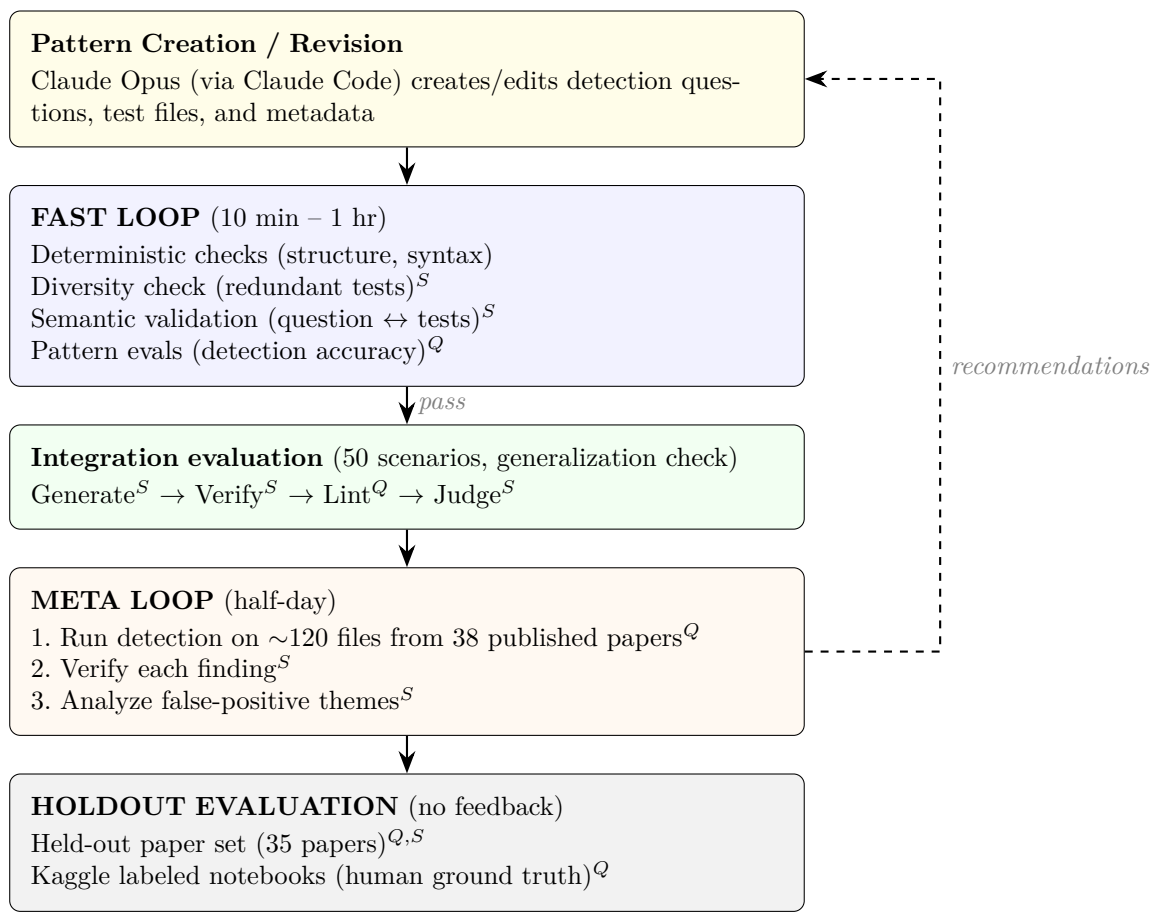
\begin{figure}[ht]
\centering
\begin{tikzpicture}[
  box/.style={draw, rounded corners, text width=10cm, align=left, inner sep=8pt, font=\small},
  arrow/.style={-{Stealth[length=3mm]}, thick},
  label/.style={font=\small\itshape, text=gray}
]

\node[box, fill=yellow!10] (create) {
  \textbf{Pattern Creation / Revision}\\[2pt]
  Claude Opus (via Claude Code) creates/edits detection questions, test files, and metadata
};

\node[box, below=0.5cm of create, fill=blue!5] (fast) {
  \textbf{FAST LOOP} (10 min -- 1 hr)\\[2pt]
  Deterministic checks (structure, syntax)\\
  Diversity check (redundant tests)$^S$\\
  Semantic validation (question $\leftrightarrow$ tests)$^S$\\
  Pattern evals (detection accuracy)$^Q$
};

\node[box, below=0.5cm of fast, fill=green!5] (integ) {
  \textbf{Integration evaluation} (50 scenarios, generalization check)\\[2pt]
  Generate$^S$ $\to$ Verify$^S$ $\to$ Lint$^Q$ $\to$ Judge$^S$
};

\node[box, below=0.5cm of integ, fill=orange!5] (meta) {
  \textbf{META LOOP} (half-day)\\[2pt]
  1.\ Run detection on $\sim$120 files from 38 published papers$^Q$\\
  2.\ Verify each finding$^S$\\
  3.\ Analyze false-positive themes$^S$
};

\draw[arrow] (create) -- (fast);
\draw[arrow] (fast) -- node[right, label] {pass} (integ);
\draw[arrow] (integ) -- (meta);

\node[box, below=0.5cm of meta, fill=gray!10] (holdout) {
  \textbf{HOLDOUT EVALUATION} (no feedback)\\[2pt]
  Held-out paper set (35 papers)$^{Q,S}$\\
  Kaggle labeled notebooks (human ground truth)$^Q$
};

\draw[arrow] (meta) -- (holdout);

\draw[arrow, dashed] (meta.east) -- ++(1.8,0) |- node[near start, right, label] {recommendations} (create.east);

\end{tikzpicture}
\caption{Self-improvement pipeline. $^S$\,Claude Sonnet 4.6, $^Q$\,Qwen3-8B (local vLLM).}
\label{fig:pipeline}
\end{figure}

Pattern development follows the layered pipeline shown in Figure~\ref{fig:pipeline}. Claude Opus (via Claude Code) creates or revises detection questions, test files, and metadata. Each pattern must pass a fast loop of quality gates (10 min -- 1 hr) before advancing to integration evaluation on 50 generated scenarios. The current 50 integration scenarios are a fixed stored set, re-evaluated after each change. Additional scenario sets could be generated to serve as a shorter feedback loop (splitting into feedback and holdout sets), or generated on the fly for fresh generalization checks at the cost of additional time and tokens.

The meta loop scans $\sim$120 self-contained files from 38 published scientific papers, verifies each finding, and runs error analysis that groups false positives into themes and produces specific recommendations for improving detection questions and test files. These recommendations feed back into pattern creation for the next cycle.

The holdout evaluations (held-out paper set, Kaggle labeled notebooks) run independently with no feedback into pattern development, providing unbiased performance estimates. A broader set of unlabeled Kaggle notebooks (also available from \citet{yang2022leakage}) and additional papers from the PapersWithCode corpus could serve as additional feedback corpora, applying the same verify-and-analyze approach.

Over three iterations of this full cycle, feedback-set precision rose from 20\% to 45\% to 62\%, driven primarily by eliminating false positives (408 $\to$ 111 $\to$ 45) while retaining most valid findings (116 $\to$ 99 $\to$ 85). For example, rep-002 (CUDA non-determinism) initially flagged all CUDA code; error analysis identified ``inference-only code'' as the dominant false-positive theme, and the detection question was refined to require evidence of training (\texttt{optimizer.step()}, \texttt{loss.backward()}). The developer orchestrates the cycle (triggering each step) but does not manually edit detection questions or test files; Claude Code applies all error-analysis recommendations and the quality gates validate the result. Each full cycle takes half a day to a full day, with local GPU time dominating; the frontier model cost is covered by a flat-rate Claude Code subscription (\$100--200/month at the time of writing), making iteration economics predictable rather than per-query.

\subsection{Design Tensions}

\textbf{Precision vs.\ recall.} For adoption, false positives are costly (noisy tools get ignored). For scientific integrity, false negatives are costly (missed data leakage can invalidate a paper). Current results reflect this tension: 75\% of papers yielded verified real bugs, but 62\% overall precision means over a third of findings are still noise.

\textbf{Test clarity vs.\ data leakage.} Test descriptions help contributors but would leak ground truth to the runtime model. Resolution: descriptions live in \texttt{pattern.toml}; test files remain pure code.

\textbf{Pattern effort vs.\ generalization.} Fewer test files means faster development; more diverse files prove the detection question captures the concept rather than overfitting. Resolution: minimum 3 positive and 3 negative per pattern, with AST similarity checks rejecting copy-paste variations.

\subsection{Scale}

The core detection implementation accounts for roughly one-seventh of the Python codebase. The rest is quality infrastructure: pattern definitions, tests, evaluation pipelines, and development tools. This ratio is deliberate: the infrastructure is what makes the tool maintainable without its original author, and what distinguishes it from thesis-scoped tools like dslinter and mllint where maintenance stopped at graduation.

\section{Deployment}

Prior ML-specific linters treated deployment as secondary to detection; as Table~\ref{tab:tools} shows, this contributed to their lack of adoption. scicode-lint treats deployment as a co-equal design requirement.

\begin{description}[style=unboxed, leftmargin=0pt, itemsep=0.5em]
\item[Individual researcher.] A workstation GPU with 20GB+ VRAM runs vLLM locally. Code never leaves the machine.
\item[Shared institutional service (recommended).] A research computing group runs a vLLM instance on a single GPU server exposing an OpenAI-compatible endpoint, similar to institutional Jupyter hubs. Any researcher uses scicode-lint from a laptop without a GPU; code stays within the institutional network.
\item[CI/CD integration.] GitHub Actions or GitLab CI can flag methodology issues before merging.
\item[AI agent feedback loop.] Structured JSON output lets AI coding agents (Claude Code, Cursor, Aider) parse findings and auto-correct methodology bugs.
\end{description}

The local-first design eliminates vendor dependency, per-query costs, and data transmission.

\section{Limitations}

\begin{description}[style=unboxed, leftmargin=0pt, itemsep=0.5em]
\item[Early-stage, single developer.] Not yet validated by external users or institutions; generalization beyond one project and one developer is unproven.
\item[Detection sensitivity to question formulation.] Detection quality depends heavily on how the detection question is formulated. Small changes in wording can shift a pattern's precision and recall substantially. The self-improvement loop provides a systematic path toward better formulations.
\item[Non-determinism.] LLM-based analysis may produce different findings on repeated runs.
\item[LLM-as-judge bias.] LLM judges may systematically over- or under-validate certain finding types, skewing precision estimates. Rotating judges across providers (OpenAI, Google, Mistral, etc.) may reduce this risk; the Kaggle evaluation (Section~4.1) provides an independent calibration point with human-labeled ground truth.
\item[Precision-only real-world evaluation.] The PapersWithCode evaluation measures precision but not recall; establishing ground truth for recall on real scientific code would require exhaustive manual review.
\item[Single-file analysis.] Many methodology issues span multiple files.
\item[Self-improvement loop tested on one corpus.] The feedback-driven pattern refinement process has been applied to one paper set; the holdout set (Section~4.2) confirms generalization (54\% vs.\ 62\% P), but the loop has not yet been validated across library version changes.
\item[Model-specific.] Results are specific to Qwen3-8B, chosen for its thinking capability (reasoning before answering improves accuracy from $\sim$78\% to $\sim$98\% on curated test files), FP8 quantization fitting 16GB VRAM, and open-source availability. No cross-model comparison was conducted.
\item[Coverage.] 66 patterns is a starting point, not comprehensive.
\end{description}

\section{Conclusion}

The detection problem is feasible; the sustainability problem is what has limited adoption of prior tools. scicode-lint's two-tier architecture addresses both by separating expensive pattern design (frontier models, run once) from cheap pattern execution (small local model, run everywhere). Maintenance costs tokens, not engineering hours.

The real-world results are encouraging but uneven. High and medium-severity patterns reach 68--72\% precision; critical findings and some leakage types remain difficult for the current single-file, 8B-parameter configuration. Multi-file context, larger runtime models, and continued pattern refinement through the self-improvement loop are concrete paths to closing these gaps. The two-tier architecture is designed to benefit from these trends automatically: as small open models grow more capable and GPUs with larger VRAM become commodity hardware, the runtime tier improves without changes to the patterns or infrastructure.

The broader point is that as AI-for-science grows and AI-generated code proliferates, the volume of scientific ML code will far outpace human review capacity. General-purpose code review tools are scaling to meet the demand for catching logic and security bugs; what remains unaddressed is methodology correctness: whether the science is right. The two-tier architecture, evaluation harness, and self-improvement process described here are domain-agnostic and open source. The tool is a starting point; the approach is transferable.

\bibliographystyle{plainnat}
\bibliography{references_v2}

\appendix
\section{Example Findings from Published Scientific Code (AI/ML for Science)}

Verified valid findings from the PapersWithCode evaluation (Section~4.2), one per category:

\textbf{Overlap leakage (ml-009).} A time series forecasting repository creates sliding-window sequences from a single series, then splits into train/val/test. With stride smaller than window size, adjacent sequences share data points, inflating evaluation metrics.

\medskip\hrule\medskip

\textbf{Loss tensor accumulation (pt-009).} A climate model emulator accumulates loss tensors, keeping the computation graph attached. Memory grows unboundedly.
\begin{lstlisting}
# Bug: graph stays attached, memory grows each iteration
self._total += nn.functional.l1_loss(pred, target)

# Fix: detach from graph
self._total += nn.functional.l1_loss(pred, target).item()
\end{lstlisting}

\medskip\hrule\medskip

\textbf{Float equality comparison (num-001).} A conditional density estimation test compares floating-point values with exact equality.
\begin{lstlisting}
# Bug: floating-point equality
assert (box_transform(x).round(2) == example_array).all()
assert np.sum(density[idx]) == 0

# Fix: tolerance-based comparison
assert np.allclose(box_transform(x).round(2), example_array)
assert np.isclose(np.sum(density[idx]), 0)
\end{lstlisting}

\medskip\hrule\medskip

\textbf{Missing map\_location (pt-015).} A surface reconstruction script loads a model checkpoint without specifying device mapping. Fails on CPU-only systems if saved on GPU.
\begin{lstlisting}
# Bug: fails if checkpoint was saved on GPU
model = torch.load(path_weight)

# Fix: explicit device mapping
model = torch.load(path_weight, map_location="cpu")
\end{lstlisting}

\medskip\hrule\medskip

\textbf{Incomplete random seeding (rep-001).} A vocoder script accepts a seed parameter but never applies it.
\begin{lstlisting}
# Bug: seed argument accepted but ignored
parser.add_argument("--random-state", type=int, default=42)
args = parser.parse_args()
# np.random.seed() never called with args.random_state

# Fix: apply the seed
np.random.seed(args.random_state)
random.seed(args.random_state)
\end{lstlisting}

\end{document}